\documentclass[twocolumn,preprintnumbers,amsmath,amssymb]{revtex4-1}
\usepackage{epsfig}
\usepackage{color}
\usepackage{amsmath}
\begin{document}
\title{The mold integration method for the calculation of the crystal-fluid interfacial free energy from simulations.}  
\author{J. R. Espinosa, C. Vega and E. Sanz}
\affiliation{Departamento de Qu\'{\i}mica F\'{\i}sica,
Facultad de Ciencias Qu\'{\i}micas, Universidad Complutense de Madrid,
28040 Madrid, Spain}

\date{\today}
\begin{abstract}

The interfacial free energy between a crystal and a fluid, $\gamma_{cf}$,  
is a highly relevant parameter in phenomena such as wetting or crystal nucleation 
and growth. Due to the difficulty of measuring $\gamma_{cf}$ experimentally,
computer simulations are often used to study the crystal-fluid interface. 
Here, we present a novel simulation methodology for the calculation of $\gamma_{cf}$. 
The methodology consists in using a mold composed
of potential energy wells to induce the formation of a crystal slab in the fluid at coexistence conditions. 
This induction is done along a reversible pathway along which the free energy difference
between the initial and the final states is obtained by means of thermodynamic integration. 
The structure of the mold is given by that of the crystal lattice planes, which allows to easily 
obtain the free energy for different crystal orientations.   
The method is validated by calculating $\gamma_{cf}$ for previously studied systems, namely the hard spheres and the Lennard-Jones
systems. Our results for the latter show that the
method is accurate enough to deal the anisotropy of $\gamma_{cf}$ with respect to the
crystal orientation. We also calculate $\gamma_{cf}$ for a recently proposed continuous version of
the hard sphere potential and obtain the same $\gamma_{cf}$ as for
the pure hard sphere system. 
The method can be implemented both in Monte Carlo and Molecular
Dynamics. In fact, we show that it can be easily used in combination with the popular Molecular Dynamics  
package GROMACS.

\end{abstract}

\maketitle


\section{Introduction}

A fluid and a crystal at coexistence are divided by a flat interface. 
The work needed to create such interface per unit area is known as the
interfacial free energy. 
The crystal-fluid interfacial free energy, $\gamma_{cf}$, is a highly relevant quantity 
due to its central role in important phenomena such as wetting or crystal nucleation 
and growth \cite{Boettinger00,Woodruff,kelton}.
Despite its importance, $\gamma_{cf}$ is still unknown for many substances
given that  there is not an easy and reliable way of measuring 
$\gamma_{cf}$ experimentally \cite{kelton2}. This difficulty contrasts with 
the determination of the fluid-fluid interfacial free energy, a task for which there are well established 
experimental and computational techniques \cite{rowlinson_widom_book,JCP_2005_123_134703}. 
Unfortunately, these techniques are not easy to implement when one of the phases involved in the coexistence has 
an infinitely large viscosity, as the crystal phase has. Moreover, $\gamma_{cf}$ 
is anisotropic and depends on the orientation of the crystal with respect to the fluid.  
The situation for water, arguably the most important substance on earth, is a good example of the difficulty 
of measuring $\gamma_{cf}$: while it is well known that the interfacial tension of liquid water
at ambient conditions is 72 mN/m, the reported values for the
ice-water interfacial free energy at ambient pressure range from to 25 to 35 mN/m \cite{pruppacher1995}.

Computer simulations can be used to assess experimental
measurements of $\gamma_{cf}$ and to improve our understanding
on the crystal-fluid interface at a molecular scale.
An important effort has been devoted to develop
simulation methodologies to calculate the crystal-fluid interfacial
free energy. 
To the best of our knowledge, these are the existing 
computational methods for the calculation of the crystal-fluid 
interfacial free energy: the cleaving method, the capillary fluctuation method, 
the methadynamics method, the tethered Monte Carlo method, and the Classical Nucleation Theory 
method.
The cleaving method, proposed by Broughton and Gilmer in 1986 \cite{broughton:5759}, was the first method devised to directly compute $\gamma_{cf}$ in a simulation. 
In this scheme the reversible work needed to cleave and re-combine the crystal and the fluid is calculated by thermodynamic
integration. 
This method is still in use and an improved version of it has been recently employed to calculate $\gamma_{cf}$ for several 
water models \cite{PhysRevLett.100.036104,doi:10.1021/ct300193e}, hard particles \cite{davidchack010,PhysRevE.74.031611}, Lennard Jones \cite{davidchack:7651}
and dipolar fluids \cite{wangmorrisJCP2013}. 
The cleaving method has recently been further improved by sorting out some
hysteresis issues \cite{horbachJCP2014}. 
In the tethered Monte Carlo scheme  \cite{verrocchioPRL2012} a complex order parameter is used to allow for a continuous transition between the fluid and the solid. 
This method has been applied to the hard sphere system \cite{verrocchioPRL2012}.
The Metadynamics method \cite{PhysRevB.81.125416} uses the rare event simulation technique Metadynamics \cite{PNAS_2002_99_12562}
to obtain the work of formation of the interface from a fluid at coexistence. 
This methodology
was originally applied to a Lennard-Jones system \cite{PhysRevB.81.125416} and has also
been used to assess experimental measurements of $\gamma_{cf}$ for Pb \cite{angiolettijpcm2011}.
The crystal-fluid interfacial free energy can also be indirectly estimated 
by combining simulation measurements of the size of critical nuclei with classical nucleation 
theory \cite{bai:124707}. This approach has been used, e. g., to 
estimate $\gamma_{cf}$ for chlatrates \cite{knott12} or water \cite{jacs2013}.

All the aforementioned methods have proved successful in the calculation of $\gamma_{cf}$ for a number of systems.
However, not all methods are equally good in terms of accuracy, simplicity and computational cost. 
The anisotropy of $\gamma_{cf}$ for different
crystal orientations is easy to study with the capillary fluctuation and with the cleaving 
methods, whereas dealing with the orientation of the crystal 
with respect to the fluid is not so trivial for the other methods. 
On the other hand, while the Metadynamics and the tethered Monte Carlo methods 
converge well for relatively small system sizes, the capillary fluctuation and the classical nucleation methods
require large system sizes. 
From a practical point of view there are methods simple to implement, like that based on 
classical nucleation theory, or more cumbersome ones like the cleaving method 
that requires following a multi-step thermodynamic route. Moreover, all methods but 
the cleaving require a local-bond order parameter in order to either detect the interface
or induce its formation. Such order parameter, used to distinguish liquid-like from 
solid-like molecules, may be difficult to conceive if the structure of the
crystal lattice is complex.
In this work we present a simple method for the direct calculation of $\gamma_{cf}$ 
that gives accurate results even for relatively small system sizes. 
The calculation of $\gamma_{cf}$ for different crystal orientations is trivial with this 
methodology. The method can be easily implemented in a bespoke Monte Carlo (MC) code 
or even in open access Molecular Dynamics (MD) packages as GROMACS \cite{hess08}. 
In brief, we use a mold of potential energy wells placed at the positions
of the atoms in a lattice plane to induce the formation of a crystal slab in the fluid at coexistence conditions. 
The work of formation of the crystal slab, obtained via thermodynamic integration, 
is directly related to the interfacial free energy.  

We test the method by calculating the interfacial free energy
of hard spheres (HS) and Lennard-Jones (LJ) (for several orientations of the
crystal) and by comparing the results with values published in the literature
\cite{JCP_2006_125_094710,Davidchack06,davidchack010,davidchack:7651,PhysRevB.81.125416}.
Moreover, we compute $\gamma_{cf}$ for the pseudo hard-sphere potential
recently proposed in Ref. \cite{pseudoHS}.

\section{The mold integration method}

\subsection{Description of the method}

\begin{figure}[h!]
\centering
\includegraphics[clip,scale=0.3]{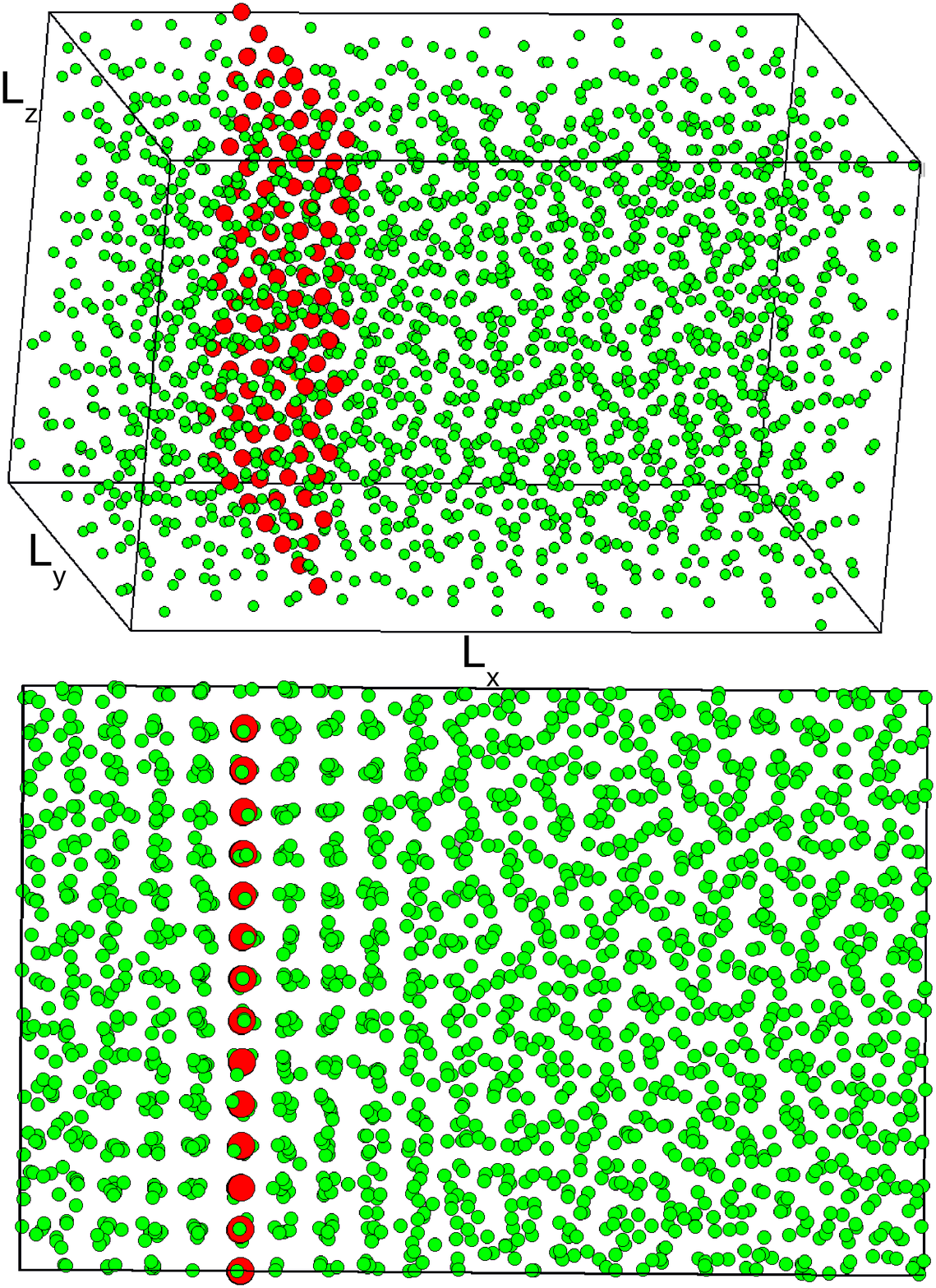}
\caption{Top: snapshot of a hard-sphere fluid at coexistence (green particles). Bottom: snapshot of a fluid with a thin crystal slab at coexistence conditions
(a projection on the
$x-z$ plane is shown). The diameter of green particles  has been reduced to 1/4 of its original size.  
The mold that induces the formation of the crystal slab is conformed by a set of potential energy wells (red spheres) whose positions
are given by the lattice sites of the selected crystal plane at coexistence conditions.
The interaction between the mold and the hard-spheres is switched off in the top configuration and on in the bottom one.   
}
\label{sketch}
\end{figure}

In this section we describe a new methodology
to compute the
interfacial free energy between a crystal and a fluid,  
$\gamma_{cf}$, by means of computer simulations. 
The basic idea is to  
reversibly
induce the formation of a thin crystalline slab
in the fluid (see Fig. \ref{sketch} for snapshots 
of a fluid and a fluid with a crystal slab).  
The work needed to form such
crystalline slab, $\Delta G^s$, is related to $\gamma_{cf}$. Because the formation of
the crystal slab is performed at coexistence conditions 
the fluid 
and the fluid plus the crystal slab have the same chemical 
potential. Then, $\Delta G^s$ is 
just the interfacial free energy times the area of the interface times 2.
The factor of 2 is due to the fact that when the crystal slab is formed 
two crystal-fluid interfaces
are created (see Fig. \ref{sketch}, bottom). 
Thus, $\gamma_{cf}$ can be simply obtained as:
\begin{equation}
\gamma_{cf} = \frac{\Delta G^s}{2A}.
\label{gamma}
\end{equation} 

In order to induce the formation of the crystal slab we use a mold composed of potential
energy wells.  The location of the wells is given by that of the particles in the lattice 
plane whose $\gamma_{cf}$ is calculated. 
In Fig.
\ref{sketch} we show a snapshot of the mold used for the calculation 
of $\gamma_{cf}$ for the 100 plane of hard spheres (red spheres).  
Each potential well must be small enough so that it can 
only accommodate one particle. 
When the mold is switched off,
particles freely diffuse in the fluid (Fig. \ref{sketch}, top). On the contrary,
when the mold is switched on, every well contains a particle and, if 
the wells are sufficiently narrow, a crystal slab
is formed at coexistence conditions (Fig. \ref{sketch}, bottom).  Typically, the mold
consists of 1 or 2 crystalline planes. In Fig. \ref{sketch} we show a mold
composed of a single plane. When filled with particles, the mold
induces the formation of crystalline planes in either side (Fig.
\ref{sketch}, bottom) thus giving rise to two crystal-fluid 
interfaces.  By gradually switching the interaction between the mold
and the particles the work of formation of the crystal slab at coexistence conditions can be
obtained by means of thermodynamic integration.   

To perform thermodynamic integration we define the following potential energy:
\begin{equation}
U(\lambda)=U_{pp}({\bf r_{1},...,r_{N}})+\lambda U_{pm}({\bf r_{1},...,r_{N};r_{w_1},...,r_{w_{N_w}}})
\label{intmold}
\end{equation}
where $N$ is the number of particles and $N_w$ the number of wells;
${\bf r_{1},...,r_{N}}$ denotes the positions of all particles and
${\bf r_{w_1},...,r_{w_{N_w}}}$ the position of the wells (which are kept 
fixed during the simulation);
$U_{pp}$ is the potential energy given by the interaction between all  
particles and $U_{pm}$ is the potential energy given by the
interaction between the mold and the particles.  $\lambda$ is a parameter that
varies from 0 to 1 connecting the initial state (mold switched off, Fig.
\ref{sketch}, top) with the final state (mold switched on, Fig. \ref{sketch},
bottom).  The interaction between the mold and the particles, $U_{pm}$, is pair
additive:
\begin{equation}
U_{pm}({\bf r_{1},...,r_{N};r_{w_1},...,r_{w_{N_w}}}) = \sum^{i=N}_{i=1} \sum^{w_j=N_w}_{w_j=1} u_{pw}(r_{iw_j}), 
\label{well-mold}
\end{equation}
where $u_{pw}(r_{iw_j})$ is a square-well interaction between the $i^{th}$ particle and
the $j^{th}$ well that depends on the distance between their centers, 
$r_{iw_j}$: 
\begin{equation}
u_{pw}(r_{iw_j})= \begin{cases} -\varepsilon, & \mbox{if }  r_{iw_j} \leq r_{w} \\ 0, & \mbox{if } r_{iw_j} > r_{w} \end{cases}.
\label{epsil}
\end{equation}
Where $r_w$ and $\varepsilon$ are the radius and the depth of the wells respectively. These are 
the only adjustable parameters of the
method. Below we explain how to deal with the tuning of these parameters. 
In any case, $r_w$ can not be larger than the particle radius to avoid
multiple filling of a single well.

By performing thermodynamic integration in $\lambda$ \cite{FrenkelSmitTI} one
can obtain the free energy difference between the fluid and the fluid plus the filled
mold, $\Delta G^m$:
\begin{eqnarray}
\Delta G^m &=& \int^{\lambda=1}_{\lambda=0} d \lambda \left<\frac{\partial U(\lambda)}{\partial \lambda}\right>_{\lambda,N,p_x,T}\nonumber\\
&=& \int^{\lambda=1}_{\lambda=0} d \lambda \left<U_{pm}\right>_{\lambda,N,p_x,T} 
\label{TI}
\end{eqnarray}
where $p_x$ and $T$ are the coexistence pressure and temperature respectively.
The mold coordinates and the edges of the simulation box
parallel to the mold are kept fixed throughout the simulation. For that purpose, the pressure is 
exerted only along the axis perpendicular to the mold, the $x$ axis in our case. Thus, 
the $x$ edge is allowed to fluctuate to keep the pressure constant, 
while the $y$ and $z$ edges do not change. The mold coordinates are not rescaled when 
a volume move is performed.   
The integrand, $\left<U_w\right>_{\lambda,N,p_x,T}$, is evaluated 
in $Np_xT$ simulations for various values of $\lambda$ and then
integrated numerically to get $\Delta G^m$.  
$\Delta G^m$ is the free energy change due to the appearance of the 
crystal slab plus that due to the interaction between the particles and the mold.  
The latter is simply given by $-N_w \varepsilon$ (recall that the well-particle interaction is
just a square well of depth $\varepsilon$). 
To calculate the interfacial free energy we are just interested in the free energy
change due to the generation of the crystal slab:
\begin{equation}
\Delta G^s = \Delta G^m + N_w \varepsilon.
\label{tie} 
\end{equation}
This equation, combined with Eq. \ref{gamma}, allows in principle for the
calculation of $\gamma_{cf}$ in a straightforward manner.  

\begin{figure}[h!] \centering
\includegraphics[clip,width=0.5\textwidth]{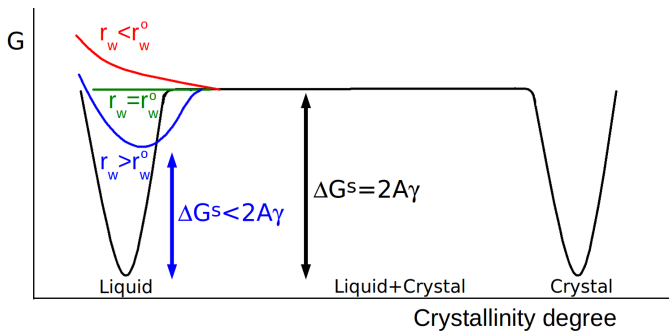} \caption{Sketch
of the free energy profile versus the crsystallinity degree for 
various potential well radius, $r_w$, at coexistence conditions. 
The black curve corresponds to the free energy profile in the absence of
any mold. The free energy is flat in between both phases because the emerging crystal slab has the same 
chemical potential as the fluid and it grows with constant crystal-fluid interfacial area.} \label{sketchfree}
\end{figure}

There is one open
issue, though: the value of $\Delta G^s$, and hence that of $\gamma_ {cf}$,
depends on $r_w$.  Therefore, one has to find a priori which value of $r_w$
gives the right value for $\gamma_ {cf}$. We shall refer to this radius as
$r_w^o$ (optimal well radius). In order to chose $r_w^o$ it is important to
understand the way in which the mold affects the free energy landscape.  In
Fig. \ref{sketchfree} we show a sketch of the free energy profile that
separates the fluid from the crystal as a function of the crystallinity degree  
(XD).  The latter can be measured, for instance,  with the aid of a
local bond order parameter that quantifies the number of crystal-like particles
in the system \cite{JCP_1996_104_09932,lechnerDellago08}.  The black curve in
Fig. \ref{sketchfree} corresponds to the free energy profile in the absence of
any mold.  The liquid and the crystal have the same free energy given that the
simulations are carried out at coexistence conditions. In between both phases
there is a free energy plateau corresponding to the presence of a crystal slab
in the fluid at coexistence conditions. Given that when a crystal slab is present there are two
interfaces of the same area $A$ (see Fig. \ref{sketch}, bottom), the free
energy difference between the plateau and the minima is $2A\gamma_{cf}$. For
$r_w > r_w^o$ the free energy profile at low crystallinity degrees changes to
that sketched by the blue curve. In this case, the free energy gap between the plateau and
the fluid's minimum is reduced by the mold, but there is still a free energy
cost to form a crystal slab. Therefore, a fluid where a mold with $r_w > r_w^o$
is switched on can remain stable for a long time before any crystal slab
arises.  If the minimum given by the blue curve is shallow (values of $r_w$
larger than $r_w^o$ but close to $r_w^o$) a fluid slab can form after some
induction period due to thermal fluctuations.  For $r_w < r_w^o$ the free
energy profile changes to that schematically shown by the red curve in Fig.
\ref{sketchfree}. Accordingly, as soon as the mold is switched on  a crystal
slab will quickly develop in order to minimize the free energy.  Therefore,
the evolution of XD depends  
on whether $r_w$ is larger or smaller than $r_w^o$.
Exploiting this difference
$r_w^o$ can be enclosed within a certain range by running simulations for
different values of $r_w$ and monitoring the behaviour of XD. 

Once $r_w^o$ is identified, one could in principle perform thermodynamic
integration for $r_w = r_w^o$ to obtain $\gamma_{cf}$.
However, in a flat landscape
as that given by $r_w = r_w^o$ (green curve in Fig. \ref{sketchfree}), XD can freely grow and may eventually 
fall into the
crystal's basin. Therefore, it is not advisable to perform thermodynamic
integration for $r_w = r_w^o$.  Instead, it is safe to integrate to states
with $r_w > r_w^o$ (blue profile in Fig. \ref{sketchfree}) for which there is a
well defined minimum in the free energy profile. Although these integrations yield 
underestimates of $\gamma_{cf}$, they provide a function $\gamma_{cf}(r_w)$
that can be extrapolated to $r_w^o$ in order to obtain the right
value of $\gamma_{cf}$. 

In summary, the method, which we call {\it mold integration method}, consists of the following steps: 
i) Preparation: The mold coordinates are obtained and an initial configuration of the fluid at coexistence conditions is prepared. The simulation box
dimensions  
must be compatible with those of the mold. 
ii) Choice of $r_w^o$: 
The mold is switched on and several simulations are run starting from the fluid configuration
previously prepared. Simulations are repeated for different values of $r_w$. 
By monitoring XD 
a range within which $r_w^o$ is enclosed is identified. iii) Calculation of $\gamma_{cf}(r_w)$:
Thermodynamic integration is performed by gradually switching the mold on.
By repeating this for several values of $r_w > r_w^o$ a 
$\gamma_{cf}(r_w)$ function is obtained. iv) The extrapolation of $\gamma_{cf}(r_w)$
to $r_w^o$ provides the definite value of $\gamma_{cf}$.   

Being the method above described completely novel, our approach
shares some features with existing methodologies. For example, in the
Metadynamics method \cite{PhysRevB.81.125416} the work needed to create a
crystal slab in a fluid at coexistence is also calculated, although in a
different way (via Metadynamics as opposed to thermodynamic integration) and
with the aid of local-bond order parameters that may be difficult to find for
complex crystal structures. This difficulty is bypassed in our method with 
the use of an {\it ad hoc} mold
In
some recent implementations of the cleaving method a cleaving potential based
on the location of the particles in the crystal plane is used
\cite{PhysRevLett.100.036104}, in resemblance to our mold of potential energy
wells.  However, whereas in our method the mold is used as a platform for the
growth of a crystal slab in the fluid, in the cleaving method the cleaving
potential is used to cleave both phases in order to subsequently recombine
them. Therefore our route to a system at coexistence is more direct than that
proposed in the cleaving framework. The use of potential wells is not exclusive
of methods for the calculation of $\gamma_{cf}$.  Potential wells have also
been used in the calculation of the free energy of amorphous and crystalline
solids via thermodynamic integration \cite{schillingJCP2009}.

\subsection{Implementation}

The implementation of the mold integration technique in MC is rather straightforward.  
A routine to evaluate the interaction between the particles and the mold, $U_{pm}$, via Eq. \ref{well-mold}
has to be incorporated to a standard $Np_xT$ MC code. $U_{pm}$ is evaluated every time a 
move is attempted and the change of $\lambda U_{pm}$
associated to the move is added to the energy change according to which 
the trial move is accepted or rejected. 
To perform thermodynamic integration via Eq. \ref{TI} the average value of 
$U_{pm}$ must be evaluated in the course of the simulation.

\begin{figure}[h!]
\centering
\includegraphics[clip,width=0.45\textwidth]{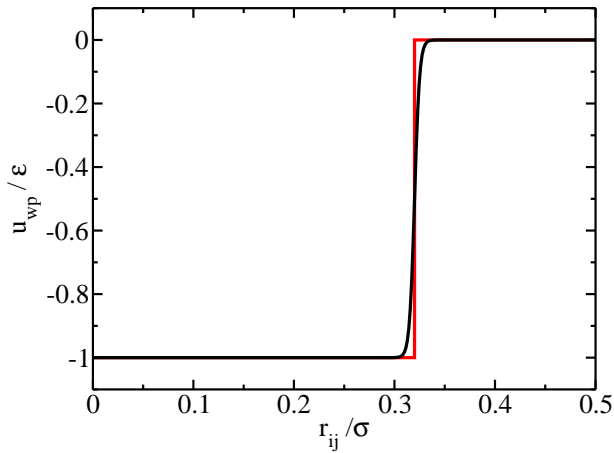}
\caption{Well-particle interaction potential for $r_w = 0.32 \sigma$. We use a square-well potential for the well-particle
interaction in MC simulations (red curve). In order to perform MD simulations we approximate the square-well interaction 
by the continuous potential given by Eq. \ref{pocilloMD} (black curve). 
In this particular example $\alpha$ in Eq. \ref{pocilloMD}  is $0.005 \sigma$.}
\label{mdsw}
\end{figure}

It is also possible to implement the mold integration technique in MD. 
We briefly discuss here how to do it for the popular MD package GROMACS \cite{hess08}. The trick
is to consider the wells as a special kind of atom. 
The interaction between the wells and the particles, Eq. \ref{epsil}, can be approximated 
by the following equation:
\begin{equation}
u_{pw}(r_{iw_j}) = -\frac{1}{2} \varepsilon \left[1 -  \quad \tanh \left(\frac{r_{iw_j}-r_w}{\alpha}\right)\right]
\label{pocilloMD}
\end{equation}
where $r_{iw_j}$, $r_w$ and $\varepsilon$ have the same meaning as in Eq. \ref{epsil} 
and $\alpha$ controls the steepness of the well's walls. This potential is continuous and differentiable
and can therefore be used in MD.  
In Fig. \ref{mdsw} we compare $u_{wp}(r_{riw_j})$ given by Eq. \ref{epsil} (black) with 
that given by Eq. \ref{pocilloMD} (red) for $\alpha = 0.005\sigma$. It is evident that a square-well interaction 
is well approximated by Eq. \ref{pocilloMD}. In GROMACS it is possible to define the 
well-particle interaction given by Eq. \ref{pocilloMD} in a tabular form, so there is no
need to modify the source code to program the interaction between the wells and the particles. 
The interaction between different wells has to be also defined in 
GROMACS in a tabular form. Such interaction is simply 0.
In order to fix the position of the wells we use the `frozen' GROMACS option. 
To perform thermodynamic integration via Eq. \ref{TI} we need to be able evaluate $<U_{pm}>$ for a given 
value of $\lambda \in [0:1]$. To do that we run the simulation with a well-particle interaction given by 
$\lambda u_{wp}$. Since GROMACS provides average values of the potential energy for any kind 
of pair interaction, one can obtain $\lambda<U_{pm}>$ as the average particle-well potential energy, 
and, in turn, $<U_{pm}>$. 
Finally, GROMACS also allows that the pressure is exerted only in one specific direction of the simulation box. 
Therefore, GROMACS includes all required tools for an easy implementation of the mold integration method in MD. 

\section{Results and discussion}

\subsection{A worked example: $\gamma_{cf}$ of hard spheres}

\subsubsection{Preparation}
The first step is obtaining the mold coordinates and a fluid configuration at coexistence conditions. 
The dimensions of the simulation box must be consistent with those 
of the mold. The mold coordinates are obtained by replicating the unit cell and taking 
a plane of the resulting lattice. 
In our case we replicate 1x7x7 times the fcc unit cell and take
a plane of atoms parallel to the $y-z$ plane. 
The resulting coordinates are shown by the red spheres in Fig. \ref{sketch}, top. 
Under periodic boundary conditions the mold is just a 100 plane of an fcc lattice. 
A configuration of the fluid at coexistence conditions (pressure = 11.54 $k_BT/\sigma^3$ \cite{Noya08})
is then prepared in a box whose $y$ and $z$ edges have the same length as the $y$ and $z$ sides of the mold. 
To achieve this we equilibrate the 
fluid in an $NpT$ simulation 
where pressure is exerted only
along the $x$ axis (we refer to this as $Np_xT$ ensemble). In this way the length of the $x$ axis, $L_x$,
is allowed to fluctuate, while $L_y$ and $L_z$ are kept fixed to the desired value
(7 times the unit cell side in this particular example).  
The resulting simulation box, that contains 1960 particles, 
is shown in Fig. \ref{sketch}, top, alongside the corresponding mold. 
We summarize the system size used for the study of the
HS system in the top row of Table \ref{phs_size}.

\begin{table}[h!]
\begin{center}
\begin{tabular}{|c|c|c|c|c|c|c|c|c|}
\hline
System & ST & $hkl$ & ($L_y$x$L_z$)/($\sigma^2$)  & $N_w$ & $N_L$ & $N$& $r_w^o/\sigma$ & $\gamma_{cf}/(\frac{k_BT}{\sigma^2})$\\
\hline
HS& MC & 100 & 10.978x10.978 & 98  & 1 & 1960 & 0.315 & 0.586 (8)\\
PHS& MD & 100 & 12.531x12.531 & 256 & 2 & 5632 & 0.375 & 0.588 (8)\\
\hline
\end{tabular}
\caption{Summary of the system size used for the calculation of $\gamma_{cf}$ for the HS and PHS models. 
ST stands for simulation type, $hkl$ for the Miller indices of the crystal plane whose $\gamma_{cf}$ is
calculated, $N_w$ for number of wells, $N$ for
number of particles and $N_L$ for number of layers (in the mold). The optimal well radius, $r_w^o$, and the
estimated value of $\gamma_{cf}$ are also reported in the table.
}
\label{phs_size}
\end{center}
\end{table}

\subsubsection{Choice of $r_w^o$}
Once the fluid is equilibrated we proceed to run $Np_xT$ simulations starting
from a fluid configuration. The mold is switched on at the beginning of the 
simulations. If the interaction between the mold and the particles is sufficiently large
all wells are quickly filled when the mold is switched on.  
We find this to be the case when $\varepsilon$ in Eq. \ref{epsil} is larger than $\sim 7 k_BT$. 
We monitor XD in the course of our simulations. As a measure of  
XD we use the following parameter, $\xi$:
\begin{equation}
\xi = \frac{\rho-\rho_f}{\rho_s-\rho_f}
\label{xi}
\end{equation}
where $\rho$ is the actual density of the system and $\rho_f$ and $\rho_s$ are the 
coexistence densities of the fluid and the solid respectively. 
Thus, $\xi$ fluctuates around 0 when the whole system is fluid, and around 1 when the whole system 
is crystalline. As a crystal slab grows in the fluid $\xi$ should
take intermediate values between the typical ones for the fluid and the crystal. 
In the appendix \ref{XD} we show that this simple 
way of quantifying XD is totally equivalent to a more sophisticated one based in 
counting the 
the number of particles in the largest cluster of solid-like particles.

\begin{figure}[h!] \centering
\includegraphics[clip,width=0.48\textwidth]{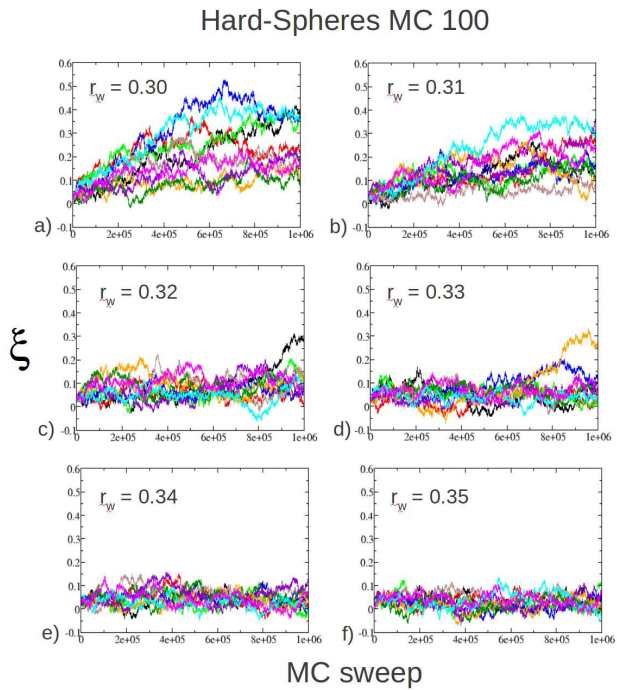} 
\caption{Crystallinity degree, XD, as measured by the parameter $\xi$ (see main text) 
as a function of simulation time for
several values of the radius of the potential wells, $r_w/\sigma$, as indicated inside
each plot. The well depth is in all cases 7.5$k_BT$. For a given $r_w$ 10 trajectories differing in the seed for 
the random number generator are started from a fluid configuration. The mold is switched on
at the beginning of the simulation. The plot corresponds to the HS potential and the 100 crystal orientation. 
} \label{widthchoice}
\end{figure}

In Fig. \ref{widthchoice} we show the evolution of $\xi$ 
for several values of $r_w$. For a given value of $r_w$ we run 10 
trajectories starting from the same initial configuration in order to 
have a statistical picture of the behaviour of the system upon
switching the mold on. The trajectories differ in the 
seed for the random number generator. 
Each $Np_xT$ MC simulation 
consists of a million sweeps.
A sweep, in turn, consists of a displacement attempt per particle plus a volume move.     
The displacement shifts for volume and 
displacement moves are tuned so that 
an average acceptance of 30-40 per cent is attained.

Three different types of behaviour can be seen when the trajectories are inspected for each $r_w$:
(a) Behaviour consistent with the presence of a deep minimum in the free energy-XD profile:
in plots e) and f) of Fig. \ref{widthchoice} XD 
stays low and fluctuates around a certain equilibrium value for all trajectories. This is consistent 
with the situation sketched by the blue curve in Fig. \ref{sketchfree}: $r_w$ is
larger than $r_w^o$ and XD 
fluctuates around the minimum given by the blue curve. 
(b) Behaviour not consistent with the presence of a minimum the free energy-XD profile: 
in plots a) and b) of Fig. \ref{widthchoice} XD 
readily grows as the mold is switched on and each trajectory evolves differently from 
the others.  
 This corresponds to
the situation illustrated by the red curve in Fig. \ref{sketchfree}: $r_w$ is
smaller than $r_w^o$ and, due to the presence of the mold, there is no free energy penalty for the 
growth of XD.
(c) Behaviour consistent with the presence of a shallow minimum in the free energy-XD profile:
In plots c) and d) of Fig. \ref{widthchoice}
XD fluctuates around an equilibrium value for 
some trajectories although, stochastically, there are trajectories
that visit high values of XD.  
For instance, for $r_w=0.33 \sigma$ the trajectory given by the orange curve
stays at low XD until it jumps around $7\cdot10^5$ MC sweeps. 
This phenomenology suggests that there is a minimum 
in the free energy profile as indicated in Fig. \ref{sketchfree} by the blue curve. 
However, the gap between the minimum and the horizontal plateau 
is not high and can be stochastically overcome by thermal activation. 

Since the optimal radius, $r_w^o$, must be in between the 
highest $r_w$ that shows no hint of a minimum ($r_w=0.31 \sigma$) and the lowest 
that does show it ($r_w=0.32 \sigma$) we take $r_w^o=0.315\pm0.005 \sigma$.

In summary, the recipe to find $r_w^o$ is to look for a value of $r_w$ that is comprised in between the 
largest one that shows no indication of the presence of a minimum in the
free energy-XD profile 
and the smallest one that does show it.
A given $r_w$ shows no indication of a minimum 
if XD can grow and evolves in a different way for different trajectories. 
By contrast, a given $r_w$ shows indication of a minimum if 
some trajectories show that XD fluctuates around a low constant value. 
In Appendix \ref{fep} we show how the free energy profile along the XD coordinate 
can be estimated for each well radius using the information contained in Fig. \ref{widthchoice}. This is quite helpful to identify which radii generate
a free energy profile with a minimum and which ones do not.

\subsubsection{Calculation of $\gamma_{cf}(r_w)$}

\begin{figure}[h!]
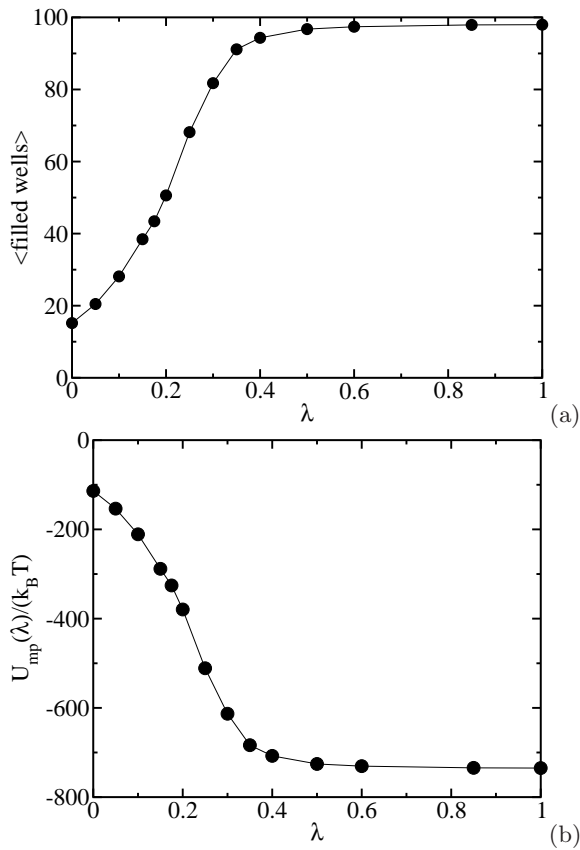
 \centering
\includegraphics[clip,width=0.40\textwidth]{FIG5A.eps}(a)
\includegraphics[clip,width=0.40\textwidth]{FIG5B.eps}(b) 
\caption{(a) average number of filled wells as a function of the parameter $\lambda$ that 
controls the strength of the interaction between the mold and the particles. (b) the 
integrand of eq. \ref{TI} is plotted against $\lambda$. both plots correspond to 
the 100 face of HS and to a 98-well mold with $r_w = 0.34 \sigma$ and $\varepsilon = 7.5 k_BT$.} \label{thermint}
\end{figure}

Once we get a value for $r_w^o$ 
we proceed to calculate $\gamma_{cf}(r_w)$ for $r_w > r_w^o$ 
by means of thermodynamic integration (Eq. \ref{TI}).
Thermodynamic integration is performed by gradually switching the mold on 
in such way that all wells are filled with particles when the 
upper integration limit is reached ({\it i.e.} when $\lambda = 1$ in Eq. \ref{TI}). 
In Fig. \ref{thermint} (a) we plot the average number of filled wells 
versus the parameter $\lambda \in [0:1]$
that controls the strength of the interaction between the 
particles and the mold via Eq. \ref{intmold}.  
Each point in Fig. \ref{thermint} is obtained in an $Np_xT$ MC simulation 
consisting of $3.3\cdot10^5$ equilibration sweeps and $6.7\cdot10^5$ production sweeps.
The plot in Fig. \ref{thermint} corresponds to a well-particle interaction parameter
$\varepsilon = 7.5~k_BT$ (Eq. \ref{epsil}) and to an $r_w = 0.34 \sigma$.
The value of $\varepsilon$ must guarantee that every well contains a particle for $\lambda = 1$. 
Provided that this condition is fulfilled, $\varepsilon$ can take any value. However, 
it is convenient that $\varepsilon$ is not too large so that the integrand varies
smoothly as $\lambda$ increases. 
In the particular case study we present here the mold is conformed by 98 wells (see Fig. \ref{sketch}, top) 
As shown in Fig. \ref{thermint} (a) all 98 wells are filled when $\lambda$ approaches 1.  
On average, about $17$ wells are occupied when the mold is switched off. 
The curve that is actually integrated in Eq. \ref{TI} is shown in Fig. \ref{thermint} (b). 
The integrand, $U_{pm}$, is simply given by the product between the average number
of filled wells and $-\varepsilon$. The integral of the curve shown in Fig. \ref{thermint} (b)
is $\Delta G^m = -600.123~k_BT$, which gives the free energy difference
between the system with the mold on and the system with the mold off. 
To simply get the free energy difference between the 
fluid and the fluid having the structure induced by the mold, $\Delta G^s$, we need to subtract 
to $\Delta G^m$ the interaction between the mold and the fluid: 
$\Delta G^s = \Delta G^m + \varepsilon N_w = -600.123 + 7.5 \cdot 98 = 134.88 k_BT$. 
$\Delta G^s$ divided by two times the area $L_yL_z$ gives an (under)estimate of the interfacial
free energy (Eq. \ref{gamma}): $\gamma_{cf}(r_w=0.34\sigma) = 0.560  k_BT/\sigma^2$. 
In this step $\gamma_{cf}$ is evaluated for some other values of $r_w > r_w^o$ in 
order to extrapolate $\gamma_{cf}(r_w)$ to $r_w^o$ in the following step.

\begin{figure}[h!] \centering
\includegraphics[clip,width=0.40\textwidth]{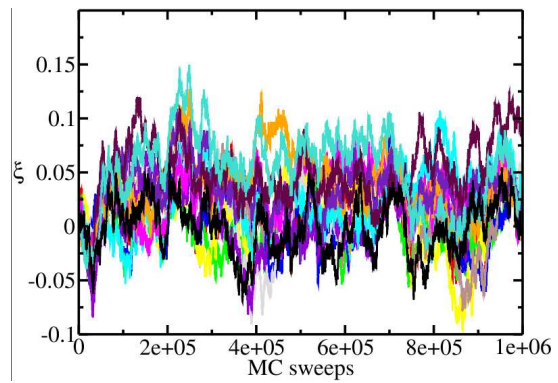}
\caption{Crystallinity degree as measured by the parameter $\xi$ for the simulations
corresponding to each integration point in Fig. \ref{thermint}. \label{check}}
\end{figure}

As previously discussed, in order for thermodynamic integration 
to be reversible we must avoid integrating at values of $r_w$ that
entail any risk that the system crystallizes. Clearly, Fig. \ref{widthchoice}
shows that such risk is negligible for $r_w = 0.34$ and $0.35 \sigma$, since XD 
fluctuates around a low, equilibrium value for all trajectories.  
The situation is not so clear for $r_w = 0.33 \sigma$, where the trajectory given by the orange curve appears to have jumped to the free energy plateau
from where the system could evolve towards the crystalline state.
Therefore, by performing thermodynamic integration at $r_w = 0.33 \sigma$ there is a small chance
that the system crystallizes in the typical simulation time required to perform thermodynamic integration.  
Hence, according to the study shown in Fig. \ref{widthchoice},  
it is safe to perform thermodynamic integration only for $r_w \ge 0.34 \sigma$. 
However, one can also try doing thermodynamic integration for $r_w$'s
closer to $r_w^o$ and validate the integration {\it a posteriori} by
checking that the system did not crystallize for any integration point. 
One of these checks is shown in Fig. \ref{check}, where we plot XD for the runs used
to compute each integration point in Fig. \ref{thermint}. XD 
stays low for all integration points, which guarantees that the integration is reversible.
It is important to do this check after performing thermodynamic integration, specially
for $r_w$'s close to $r_w^o$. 

\subsubsection{Extrapolation of $\gamma_{cf}(r_w)$}

\begin{figure}[h!]
\centering
\includegraphics[clip,width=0.45\textwidth]{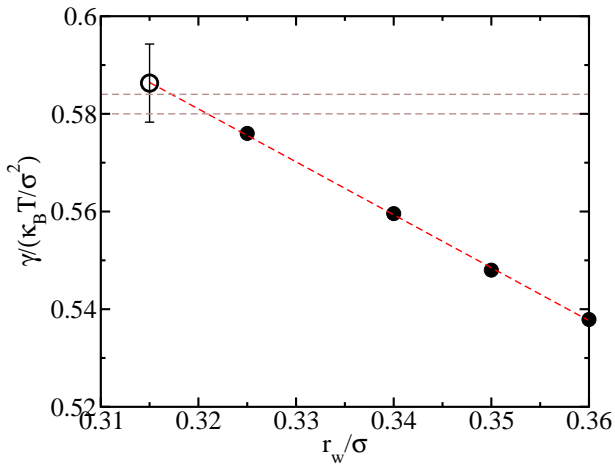}
\caption{Solid symbols: interfacial free energy versus the radius of the potential wells
for the 100 face of HS. 
The red dashed line is a linear fit to 
our data. The value of the fit at $r_w=r_w^o=0.315\sigma$ gives our estimate for the
interfacial free energy $\gamma_{cf} = 0.586~k_BT/\sigma^2$, indicated by the open symbol.
The horizontal dashed lines indicate the value
of $\gamma_{cf}$ given in Ref. \cite{davidchack010}. 
}
\label{gammaHS}
\end{figure}

Above we have discussed in detail the calculation of $\gamma_{cf}(r_w)$ for
$r_w = 0.34\sigma$. The same calculation has to be repeated for other values of 
$r_w$ in order to get a function $\gamma_{cf}(r_w)$ that can be extrapolated 
to the radius previously 
identified as the optimal one: $r_w^o = 0.315\sigma$. 
In Fig. \ref{gammaHS} we show $\gamma_{cf}(r_w)$ for the HS system. 
The dependency of $\gamma_{cf}$ on $r_w$ looks rather linear, which 
allows to easily extrapolate $\gamma_{cf}(r_w)$ to $r_w^o$. 
The extrapolation is given by the open symbol with the error bar in Fig. \ref{gammaHS}. Thus, 
our estimated value for $\gamma_{cf}$ for the 100 plane of HS is 
$\gamma_{cf} = 0.586(8) k_BT/\sigma^2$. 
The main error source in our calculation comes from the uncertainty in determining $r_w^o$. 
The uncertainty in the thermodynamic integration also contributes, although to a lesser extent,  
to our final error bar.  
Our value is in very good agreement with the most recent 
estimate of $\gamma_{cf} = 0.582(2) k_BT/\sigma^2$ \cite{davidchack010} 
(horizontal dashed lines in Fig. \ref{gammaHS}), obtained via 
the cleaving method \cite{broughton:5759}. 
Our value is also in agreement with the latest estimate of $\gamma_{cf}$ for the 100 plane of HS from capillary wave 
fluctuations: $0.57(2) kT/\sigma^2$ \cite{JCP_2006_125_094710}.

This excellent result proves the ability of our mold integration method
to evaluate the crystal-fluid interfacial free energy. The method is simple conceptually 
and easy to implement. With a 10 processors machine and our bespoke MC algorithm
the calculation of the crystal-fluid interfacial free energy for the 100 plane of HS took us about two days. 
To further validate the methodology 
in the following section we show results for the LJ system, for which we compute the
interfacial free energy not only for the 100 plane but also for the 111 plane.

\subsection{$\gamma_{cf}$ for the LJ system}
\label{LJ}

In this section we report the calculation of $\gamma_{cf}$ for the LJ model as
modified by Broughton and Gilmer \cite{broughtongilmerLJpotential1983}. We
perform the calculation at the triple point of the model, which was determined
in Ref. \cite{broughton1986} to be at $T=0.617 \epsilon /k_B$, corresponding to
a pressure $p = -0.02 \epsilon/\sigma^3$ ($\epsilon$ and $\sigma$ are the
interaction parameters of the LJ system \cite{broughtongilmerLJpotential1983}).
At these thermodynamic conditions the density of the fluid is 0.828 $\sigma^{-3}$
and that of the crystal 0.945 $\sigma^{-3}$ \cite{davidchack:7651}. 

\begin{figure}[h!]
\centering
\includegraphics[clip,scale=0.2]{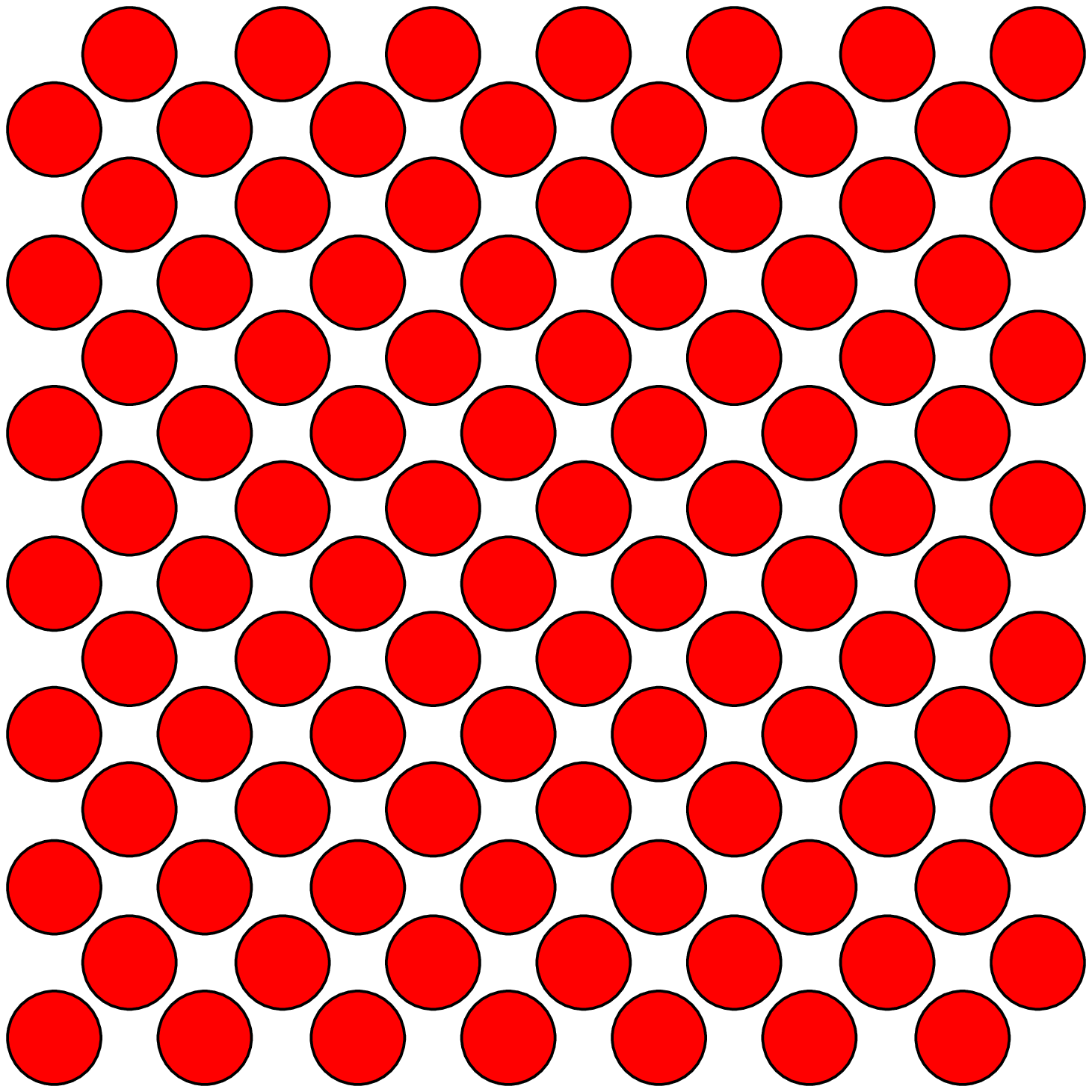}
\includegraphics[clip,scale=0.2]{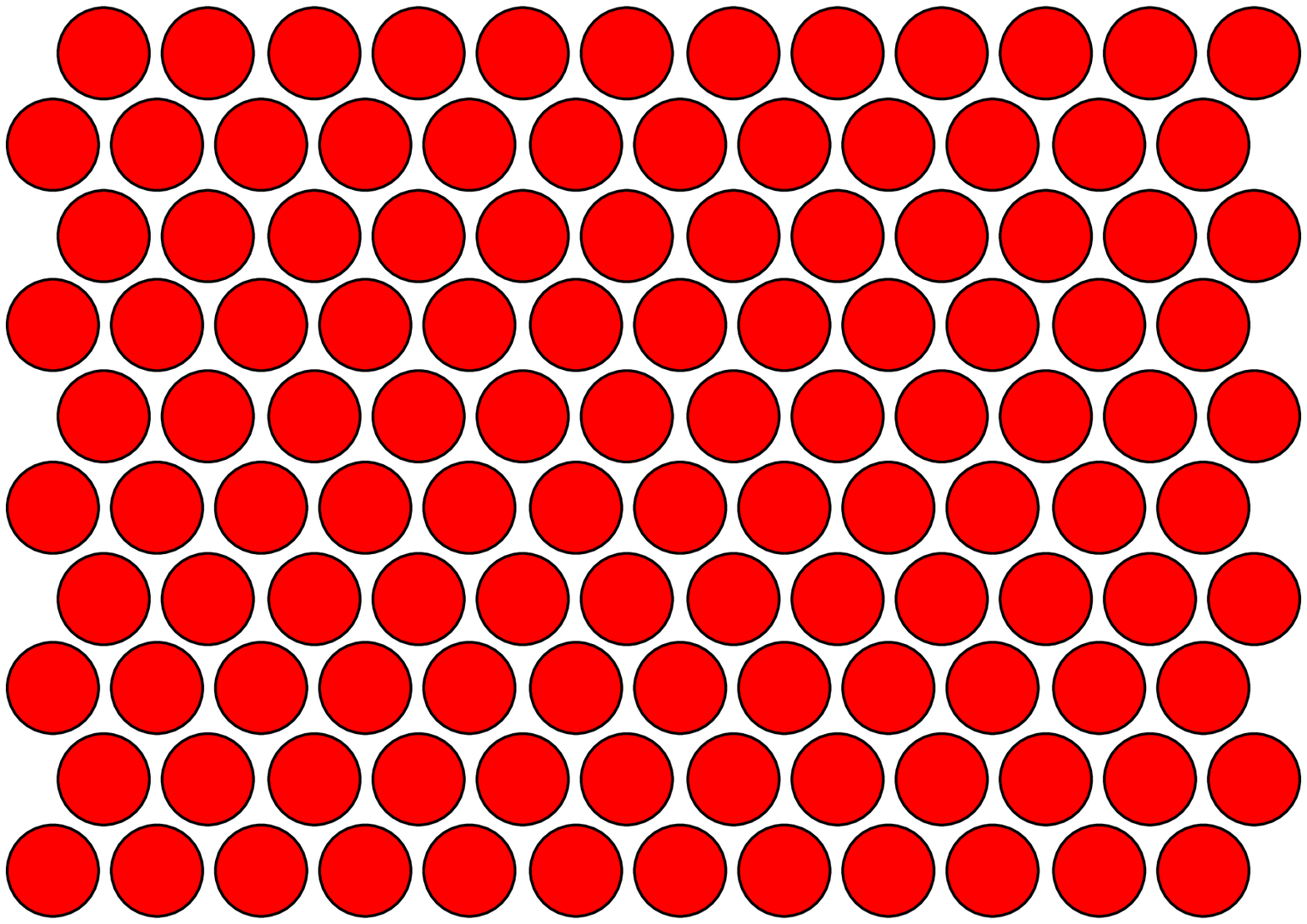}
\caption{Molds used for the calculation of $\gamma_{cf}$ for the 100 (left) and the 111 (right) crystal
orientations of the LJ system. Note the more compact packing of wells in the 111 mold.}
\label{molds}
\end{figure}

To illustrate the suitability of our methodology to deal with the anisotropy of the 
crystal-fluid interfacial free energy we calculate  
$\gamma_{cf}$ for two different crystal orientations. The orientation of the  
crystal with respect to the fluid is indicated by the Miller indices 
of the plane parallel to the interface. In this work we calculate $\gamma_{cf}$ for the
100 and the 111 orientations. 
Obtaining $\gamma_{cf}$ for a given   
crystal orientation with the mold integration method just requires using a mold coming from the
lattice plane that defines such orientation.
In Fig. \ref{molds} we show
the molds used for the calculation of the 100 (left) and the 111 (right) interfacial
free energies. 

In the previous section, where we describe the calculation of $\gamma_{cf}$ for the HS
system, we use a single layer mold. However, the mold
can be composed of more than a single layer. 
We prove in this section that using molds composed of two 
layers (bilayer mold) one obtains results which are consistent with those obtained
by using a single layer (monolayer mold). 
Moreover, we also compare in this section MC with MD. In the implementation section 
above we describe the way the mold integration method can be easily implemented in 
the popular MD simulation package GROMACS, which we use to calculate $\gamma_{cf}$
for the LJ system. 

The first step is to prepare the mold and a fluid configuration at equilibrium in a simulation 
box compatible with the mold. We give details on how to do this for the HS system in the previous 
section.   
In Table \ref{sizeLJ} we list the simulations used for the calculation of $\gamma_{cf}$
for the 100 and 111 crystal orientations for the LJ system. We indicate the area of the simulation 
box side parallel to the mold ($L_y$x$L_z$), the number of wells that conform the mold, 
the number of layers of the mold and the number of particles of the system.  

Once the initial set up is ready we run a number of trajectories (about 10) 
for different values of $r_w$ in order to find $r_w^o$. 
As previously 
described for the HS system, $r_w^o$ can be found
by looking at the behaviour of XD as a function 
of the simulation time for different $r_w$'s. 
The values of $r_w^o$ thus obtained are shown in Table \ref{sizeLJ}. 
It is interesting to realize that $r_w^o$ changes from one crystal orientation to another. 
$r_w^o$ is always smaller for the 111 plane. This shows that it is necessary to adjust 
the value of $r_w^o$ for every crystal orientation separately. 
We also show in Table \ref{sizeLJ} that $r_w^o$ also depends on the number of 
layers that conform the mold. Monolayer molds need smaller values of $r_w^o$ to induce 
the formation of a crystal slab than bilayer ones. 
The work required to fill a well increases as its radius decreases since the smaller the well's volume
the more unfavourable it becomes confining a particle inside. 
Therefore, one has to supply more energy per well to a monolayer mold in order to get the same energy per unit area as in
a bilayer mold. This seems reasonable since a bilayer mold has twice as many wells per unit area. 
Finally, from our analysis it also turns out that MD $r_w^o$s are slightly larger than MC ones 
(when compared for the same crystal orientation and the same number of layers in the mold).
This may be due to the fact that, although the continuous potential given by Eq. \ref{pocilloMD} closely follows a square-well interaction (see Fig. \ref{mdsw}), 
the equivalence is not perfect. Nevertheless, as we show below, the small difference in $r_w^o$ between MC and MD is not
reflected in the estimated value of $\gamma_{cf}$.

\begin{table}[h!]
\begin{center}
\begin{tabular}{|c|c|c|c|c|c|c|c|}
\hline
ST & $hkl$ & ($L_y$x$L_z$)/($\sigma^2$)  & $N_w$ & $N_L$ & $N$& $r_w^o/\sigma$ & $\gamma_{cf}/(\frac{\epsilon}{\sigma^2})$\\
\hline
MC & 100 & 11.323x11.323 & 98 & 1 & 1960 & 0.305& 0.372(8)\\
MD & 100 & 11.323x11.323& 98 & 1 & 1960 & 0.315& 0.372(8)\\
MD & 100 & 14.543x14.543&  324& 2 & 6480& 0.385& 0.373(8)\\
\hline
MC & 111 & 13.726x9.906 & 120 & 1 & 2160 & 0.285& 0.350(8)\\
MD & 111 & 13.726x9.906& 120& 1 & 2160 & 0.295&0.354(8)\\
MD & 111 & 13.726x9.906& 240& 2 & 2160 & 0.385&0.348(8)\\
\hline
\hline
&  100 & & & &12158& & 0.371(3) \cite{davidchack:7651} \\
&  100 & & & & 38740 & & 0.369(8) \cite{morris:3920} \\
&  100 & & & &2352& & 0.370(2) \cite{PhysRevB.81.125416} \\
&  100 & & & &1790& & 0.34(2) \cite{Broughton86} \\
\hline
&  111 & & & &8916& & 0.347(3) \cite{davidchack:7651} \\
&  111 & & & & 38740 & & 0.355(8) \cite{morris:3920} \\
&  111 & & & &1674& & 0.35(2) \cite{Broughton86} \\
\hline
\end{tabular}
\caption{Summary of the size of the systems used for the calculation of $\gamma_{cf}$ for the LJ
system. 
The meaning of ST, $hkl$, $N_w$, $N_L$ and $N$
is the same as in Table \ref{phs_size}.
For comparison, in the bottom frame of the table we give
values for $\gamma_{cf}$ obtained in previous works, alongside the number of particles used for their calculation.}
\label{sizeLJ}
\end{center}
\end{table}

\begin{figure}[h!]
\centering
\includegraphics[clip,scale=0.3]{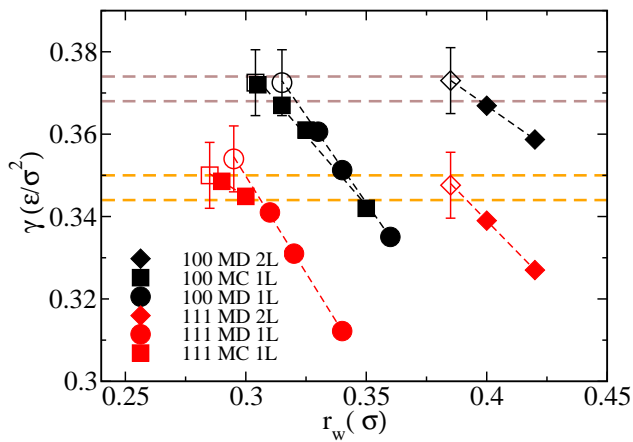}
\caption{Interfacial free energy as a function of the well radius for the LJ-like potential
proposed in Ref. \cite{broughtongilmerLJpotential1983}. 
Results are shown for two different crystal orientations (111, red symbols, and 100, black symbols),
two simulation techniques (MC, squares, and MD, other symbols), and two thicknesses of the mold
(bilayer, 2L, diamonds, and monolayer, 1L, other symbols). Filled symbols are simulation data and empty symbols
with error bars correspond to the extrapolation to $r_w^o$. Brown and orange dashed horizontal lines correspond
to the value of $\gamma_cf$ reported in Ref. \cite{davidchack:7651} for the 100 and 111 orientations respectively. Black and red dashed
lines are linear fits to the different sets of solid symbols.} 
\label{figura_magica}
\end{figure}

Once $r_w^o$ is identified for each system the next step is to 
perform thermodynamic integration for at least a couple of 
values of $r_w > r_w^o$ in order to obtain a function 
$\gamma_{cf}(r_w)$ that can be extrapolated to $r_w^o$. 
In Fig. \ref{figura_magica} we show $\gamma_{cf}(r_w)$ for all systems investigated.  
Red symbols correspond to the results for the 100 orientation and black ones to 
the 111 orientation.
Let us start by discussing the results for the 100 orientation. 
Filled symbols correspond to the calculation of $\gamma_{cf}(r_w)$ via
thermodynamic integration and empty ones to the extrapolation of 
$\gamma_{cf}(r_w)$ to $r_w^o$. The black squares correspond to MC and the  
black circles to MD simulations, both with a monolayer mold. It is clear from Fig. \ref{figura_magica} that
both MC and MD yield consistent results for the calculation of $\gamma_{cf}(r_w)$ via 
thermodynamic integration. A linear extrapolation of  
the MC and MD data to their corresponding values of $r_w^o$ 
(see Table \ref{sizeLJ}) provides an estimate for $\gamma_{cf}$, indicated by the 
open symbols in Fig. \ref{figura_magica} and 
reported in Table \ref{sizeLJ}.
Within the error of the method both MC and MD give the same $\gamma_{cf}$ for the 100 orientation.    
By using a bilayer mold (black diamonds) we get also, within error, the same value for $\gamma_{cf}$. 
Red squares and circles correspond to the results for a 111 monolayer mold
as obtained from MC and MD respectively. Again, a good agreement between both simulation techniques is obtained. 
Moreover, the results for the bilayer (red diamonds) give the same $\gamma_{cf}$ as the monolayer mold.
In summary, in Fig. \ref{figura_magica} we show that the mold integration technique gives 
consistent results regardless the simulation technique (MC or MD), or the number of layers in the mold (1 or 2). 
The accuracy of the technique is sufficient to distinguish between the interfacial free energy of two different
crystal orientations (100 and 111). 

As discussed above, in order to compute $\gamma_{cf}$ we recommend to 
obtain first two or three under-estimates of $\gamma_{cf}$ for $r_w > r_w^o$, 
where thermodynamic integration is reversible, and then extrapolate 
the results to $r_w^o$. A close inspection of Fig. \ref{figura_magica} 
shows that the MC estimate of $\gamma_{cf}$ for the 100 orientation was directly performed 
at $r_w=r_w^o$. This allows to directly estimate $\gamma_{cf}$ without 
the need of any extrapolation, but apparently contradicts the advice
of performing thermodynamic integration for $r_w > r_w^o$. 
In fact, we had to resort to a tailored type of MC move in order to 
perform thermodynamic integration for $r_w=r_w^o$, where the reversibility of
thermodynamic integration is compromised by the possibility that
the system fully crystallizes. 
Such move consisted in performing blocks of thousands
of MC sweeps that are accepted or rejected according to the
final value of XD.  
If XD increases beyond the point at which 
the system is committed
to fully crystallize, the whole block of MC sweeps is rejected and 
re-started with a different seed for the random number generator. 
Otherwise, the MC simulation continues normally. In order
to set both the length of the simulation blocks and the XD  
threshold it is necessary to get some experience first by examining 
several unbiased runs at $r_w=r_w^o$. 
In this way we have two MC estimates of $\gamma_{cf}$ for the 100 interface: 
one coming from the `direct' calculation of $\gamma_{cf}$ at 
$r_w=r_w^o$ (as described in this paragraph), and another coming from the extrapolation 
of estimates for $r_w>r_w^o$. As shown in Fig. \ref{figura_magica} both estimates
coincide pretty well, which gives us confidence in the extrapolation procedure
described in previous sections. Although both ways of estimating $\gamma_{cf}(r_w^o)$
are equally valid, we recommend the use of the extrapolation method because 
it is more general as it does not require the implementation of the MC-block
moves described in this paragraph. 

In Table \ref{sizeLJ} we show that one can obtain consistent results for molds with one 
or two layers. In principle any number of layers can be used. However, 
one must take into account that $r_w^o$ increases as the number of layers
increases (see table \ref{sizeLJ}) and that $r_w^o$ can not be larger than 0.5 $\sigma$ in order
to avoid multiple filling of the wells. For this reason, in practise,  
we could not use a mold with more than two layers to compute $\gamma_{cf}$. 
In any case, there is no practical advantage in using bi-layer over mono-layer molds. 
In fact, with a mono-layer mold the number of well-particle interactions is half as many 
and the code runs faster.

The interfacial free energy for the LJ model at the triple point has been
directly determined by Broughton and Gilmer in 1986 using the cleaving method
\cite{Broughton86}, by Laird and Davidchack using a more accurate variant of the same
methodology \cite{davidchack:7651}, by Morris
and Song using a capillary fluctuation approach \cite{morris:3920} and by Angioletti-Uberti {\it et al.}
using a Metadynamics-based approach \cite{PhysRevB.81.125416}. 
In the bottom part of table \ref{sizeLJ} we report the values of $\gamma_{cf}$ obtained in these works. 
The agreement between our data and those obtained in Refs. \cite{davidchack:7651,morris:3920} is very good.
The comparison with Ref. \cite{Broughton86} is not bad either, particularly taking into account
that in 1986 the computational resources did not allow for accurate enough calculations to distinguish between different 
crystal orientations. 
In Ref. \cite{bai:124707} $\gamma_{cf}$ was indirectly estimated via classical nucleation theory 
($\gamma_{cf} = 0.302(2) \epsilon/k_BT$). 
The discrepancy with our data may be partly due to the fact that in Ref. \cite{bai:124707}
a value of $\gamma_{cf}$ averaged over several crystal orientations is provided. 
In summary, in Table \ref{sizeLJ} we show that the values obtained from our method are in good agreement with the general consensus
reached for the $\gamma_{cf}$ of LJ for two different crystal orientations.

The extensive work done on the LJ system allows for a comparison between different simulation methods. 
In table \ref{sizeLJ} we report the number of molecules used for the calculation of $\gamma_{cf}$ for 
each simulation method. The number we report for Ref. \cite{morris:3920}, where the capillary fluctuation method was used, 
is an average over all systems employed. In terms of computational cost, the capillary fluctuation method is 
expensive. Moreover it requires
the evaluation of the spectrum of capillary waves for at least three different crystal orientations in order to 
provide a value of $\gamma_{cf}$.  
The method based on Classical Nucleation Theory also requires a large number of particles because
such theory works best for large cluster sizes \cite{bai:124707}. 
The cleaving method used in Ref. \cite{Broughton86} used 
relatively small systems, but the results were not accurate enough to distinguish between different 
crystal orientations. Many more particles were used in Ref. \cite{davidchack:7651} in order to 
gain accuracy. Both the Metadynamics method and our mold integration method 
are capable of producing accurate results for systems of less than 2000 particles. 
We have checked that there are no significant system size effects present in our simulations. In Table \ref{sizeLJ} we 
show that a calculation with 6480 particles gives the same result as one with 1960. 
Therefore, the possibility  of using small systems is a positive aspect of our method. 

Another advantage is the simplicity with which it deals with different crystal orientations 
(one simply has to use a mold coming from the corresponding lattice plane). 
Also the cleaving and the capillary fluctuation methods easily deal with 
the anisotropy of $\gamma_{cf}$. In both methods the fluid is brought into contact with the crystal 
at the desired orientation. For the capillary fluctuation method the problem is not so straightforward, though. 
What is obtained from the analysis of the capillary waves spectrum is the interfacial 
stiffness, and several orientations must be combined with a cubic harmonic expansion in 
order to obtain estimates of $\gamma_{cf}$. 
The way to deal with the anisotropy of $\gamma_{cf}$ is more complex for the other methods. 
In the Metadynamics method, for instance, an order parameter has to be
devised in order to induce the growth of the crystal. Finding a good 
order parameter may be a non-trivial task for crystal structures whose complexity 
goes beyond that of simple fcc or bcc lattices \cite{santiso:064109}. 
The classical nucleation method also requires 
an order parameter to measure the size of the embedded clusters. 

\subsection{$\gamma_{cf}$ for the pseudo hard-sphere potential}

A system composed of hard spheres (HS) is arguably the simplest non-trivial   
model having fluid, crystal and glass  phases \cite{alder:1208,N_1986_320_340}.
Therefore, this model is widely used by researchers on a diverse range of problems
like the glass transition \cite{S_2002_296_104}, crystal nucleation \cite{Nature_2001_409_1020}, 
or granular matter \cite{song2008aphase}.  
There is great interest in finding a continuous potential whose kinetic and thermodynamic behaviour  
reproduces that of the discontinuous potential of pure HS. Finding such 
potential would allow to explore the physics of the HS system with simple MD simulations. 
This is important because MD simulation packages like GROMACS are nowadays accessible 
to a large scientific community. 
For instance, having a continuous version of the HS potential would be of great 
help for the investigation 
the crystallization of hard spheres, where large discrepancies between
experimental and simulation measures of the nucleation rate have been reported (see Ref. \cite{filionjcp2011} and
references therein). 

Quite recently, Jover {\it et al.} have proposed a continuous potential 
that at reduced temperature 1.5 behaves very much alike a system of pure
HS in terms of the equation of state and the diffusion coefficient \cite{pseudoHS}.
We refer to the potential proposed by Jover {\it et al.} at reduced temperature 1.5
as the pseudo hard-sphere potential, PHS.  
Later on, Espinosa {\it et al.} showed that the coexistence pressure and densities
for the PHS model are also very similar to those of the pure HS system \cite{sigmoide}. 

The PHS potential is a good candidate for the investigation of
the crystal nucleation rate of HS given that, as mentioned before, both models have a very similar
thermodynamic coexistence, diffusion coefficient, and equation of state. 
Nothing is known, however, about the $\gamma_{cf}$ of the PHS potential, a crucial parameter
in crystal nucleation \cite{kelton}. If $\gamma_{cf}$ was also similar
to that of pure HS then the PHS model could be reliably used in MD simulations to 
obtain predictions about the crystallization behaviour of HS. Here, we use the mold 
integration method to obtain $\gamma_{cf}$ for the PHS model. 

We evaluate the interfacial free energy for the 100 crystal orientation 
in order to compare with our results for the pure HS model. We use a two-layer mold with 
128 particles in each layer. The system size is summarized in the bottom row of table \ref{phs_size}.
All simulations are performed with the GROMACS MD package.

\begin{figure}[h!]
\centering
\includegraphics[clip,width=0.48\textwidth]{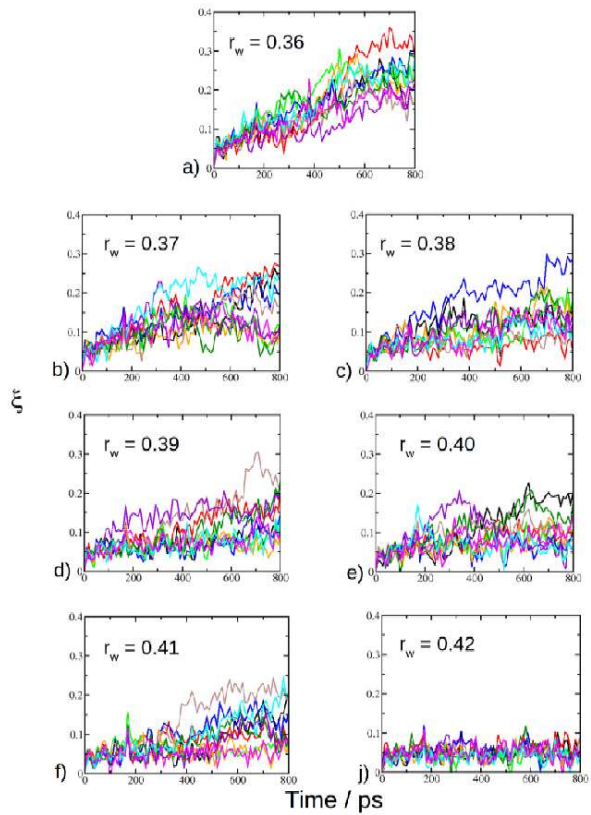}
\caption{
Crystallinity degree, XD, as measured by the parameter $\xi$ (see main text)
as a function of simulation time for
several values of the radius of the mold potential wells, $r_w/\sigma$, as indicated inside
each plot. The well depth is in all cases 7.5$k_BT$. For a given $r_w$ 10 trajectories differing in the seed for
the random number generator are started from a fluid configuration. The mold is switched on
at the beginning of the simulation. The plot corresponds to the PHS potential and the 100 crystal orientation. 
}
\label{Quasi_hard_spheres_100}
\end{figure}

To determine $r_w^o$ we monitor the XD as a function of time for several trajectories
and for several $r_w$'s. To monitor XD we use the parameter $\xi$ defined in Eq. \ref{xi}. 
The results are shown in Fig. \ref{Quasi_hard_spheres_100}. 
As explained above for the HS and LJ cases $r_w^o$ will be comprised in between 
the largest value of $r_w$ that shows no indication of the presence of a minimum in the
free energy-XD profile and the smallest $r_w$ that does show it. 
In Fig. \ref{Quasi_hard_spheres_100} $r_w/\sigma = 0.42$ clearly corresponds to the presence of a 
deep minimum (one that can not be overcome by thermal activation for any of the 10 trajectories).  
By contrast, $r_w/\sigma =$ 0.36 and 0.37 show no presence of a minimum because XD can grow and evolves
in a different way for different trajectories. 
$r_w/\sigma = 0.38-41$ correspond to a shallow minimum because in all cases there are trajectories
for which XD fluctuates for a significant period around typical values for
the deep minimum case (XD$\approx 0.05$).  
Therefore we set $r_w^o/\sigma = 0.375(5)$.

\begin{figure}[h!]
\centering
\includegraphics[clip,scale=0.3]{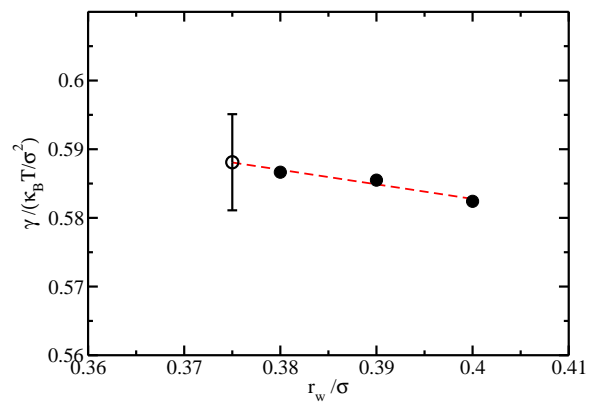}
\caption{Solid symbols: interfacial free energy versus the well radius for the PHS model. Dashed line: linear fit to the solid symbols. 
Empty symbol: extrapolation of the linear fit to $r_w^o$.}

\label{monocapa_1_0_0__hard_spheres_md}
\end{figure}

Once we get $r_w^o$ we perform thermodynamic integration for values of $r_w > r_w^o$ and obtain the 
solid points shown in Fig. \ref{monocapa_1_0_0__hard_spheres_md}. The extrapolation of these data to $r_w^o$
gives $\gamma_{cf} = 0.588(8) k_BT/\sigma^2$ that is, within the error bar, the same value we find for the pure
HS system. This result implies that the PHS model can be used with confidence for the 
study of  the behaviour of HS. The simulation details and results for the HS and PHS models are compared in Table \ref{phs_size}.

\section{Summary and conclusions}

We propose a novel simulation methodology for the calculation of the crystal-fluid interfacial 
free energy. The main idea of the method is the use of a mold of potential energy
wells to induce the formation of a crystal slab in a fluid at coexistence conditions. 
The coordinates of the mold's wells are given by the lattice positions of the crystal
plane whose interfacial free energy is evaluated.
The interaction between the wells and the fluid particles is square-well like. 
The free energy difference between the fluid and the fluid having the crystal slab
induced by the mold is obtained by means of thermodynamic integration along a reversible
path in which the wells are gradually filled. The method consists in four basic steps: (i) preparation 
of the mold and of the initial configuration of the fluid; (ii) estimation of the
optimal radius of the wells; (iii) calculation of the interfacial free energy as a function 
of the well radius by thermodynamic integration; (iv) extrapolation of the
function obtained in step (iii) to the optimal radius to get the final estimate
of the interfacial free energy. 

We validate our methodology by calculating the interfacial free energy of systems 
composed of hard spheres and Lennard-Jones particles. In both cases we find a very good agreement 
with previous estimates. Moreover, we show that our methodology is accurate enough
to discriminate between different crystal orientations of the Lennard-Jones system. 
We also use the new method to calculate the interfacial free energy 
for a continuous version of the hard sphere model for which, to the best of our knowledge, the interfacial
free energy had not been previously calculated. Within the statistical uncertainty
of our calculations we obtain the same interfacial free energy for both the continuous and the discontinuous
potentials.  

One of the main advantanges of our method with respect to existing ones
is that no local-bond order parameter is required to either detect the interface or 
induce the growth of the solid \cite{PhysRevLett.86.5530,PhysRevB.81.125416,bai:124707,verrocchioPRL2012}.
The cleaving method does not require an order parameter either \cite{Broughton86}, but it entails following a rather cumbersome 
thermodynamic route.
Our method, gives accurate results even for relatively small systems (about 2000 particles), which can not be achieved
with the capillary fluctuation \cite{PhysRevLett.86.5530} or the classical nucleation \cite{bai:124707} methods. 
Moreover, it can potentially deal with complex crystal lattices with no
extra methodological complexity.  
The method can be easily implemented either in Monte Carlo or in
standard Molecular Dynamics packages such
as GROMACS. Therefore, we hope it will be appealing to the scientific community 
interested in investigating the properties of the crystal-melt interface.

{\bf Acknowledgements}\\
All authors thank the Spanish Ministry of Economy and Competitiviness for the financial support 
through the project FIS2013-43209-P. 
E. Sanz and J. R. Espinosa acknowledge financial support from the EU grant 322326-COSAAC-FP7-PEOPLE-2012-CIG
and E. Sanz from a Spanish grant Ramon y Cajal.

\clearpage


\clearpage
\appendix

\section{Measure of XD}
\label{XD}
\begin{figure} \centering
\includegraphics[clip,width=0.48\textwidth]{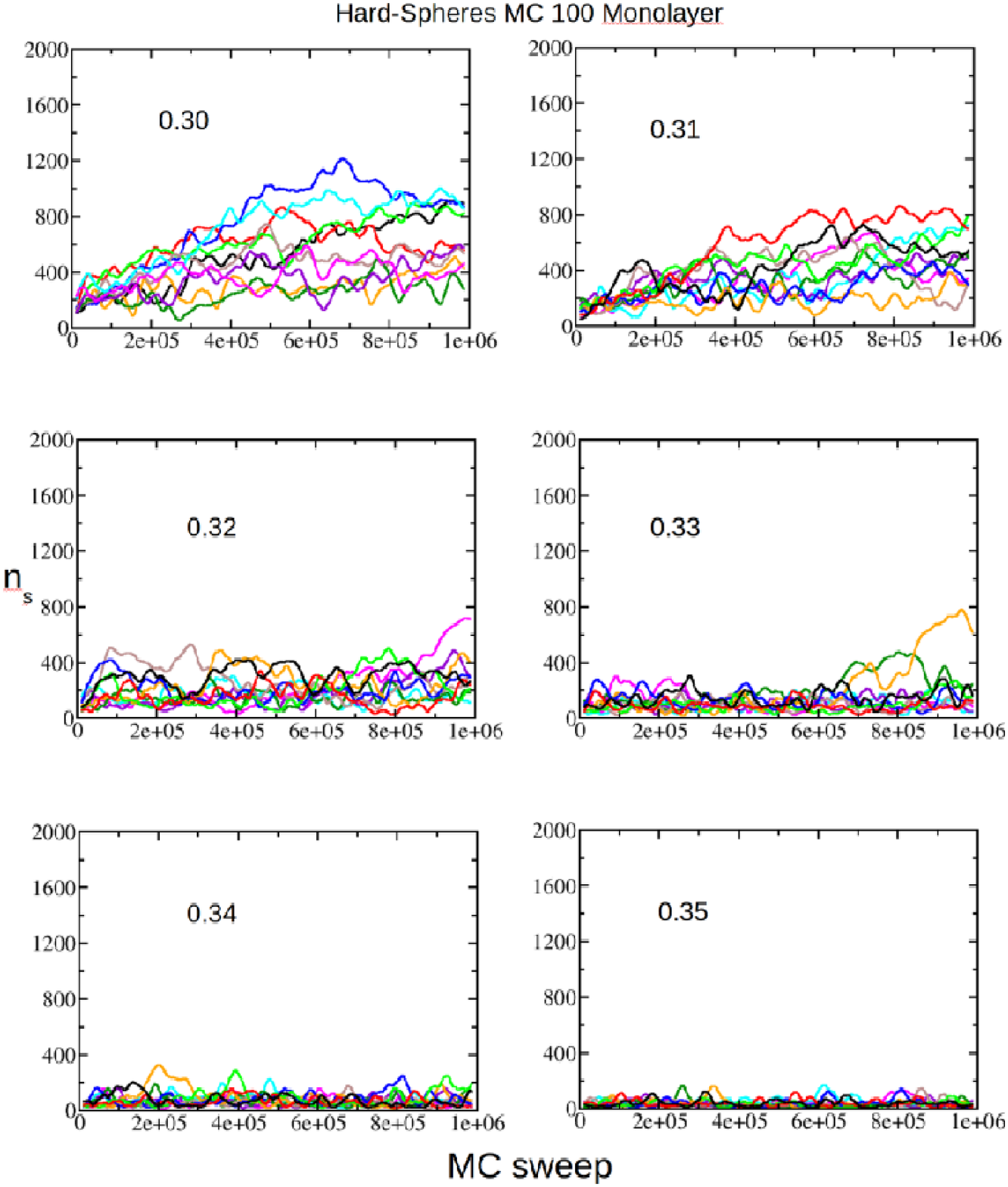} 
\caption{Crystallinity degree as measured by the number of particles in the biggest
cluster of solid-like particles, $n_s$ 
as a function of simulation time for
several values of the radius of the potential wells, $r_w/\sigma$, as indicated inside
each plot. The well depth is in all cases 7.5$k_BT$. For a given $r_w$ 10 trajectories differing in the seed for 
the random number generator are started from a fluid configuration. The mold is switched on
at the beginning of the simulation. The plot corresponds to the HS system and the 100 crystal orientation.  
} \label{widthchoicei_app}
\end{figure}

Here we show that the simple order parameter $\xi$
used to quantify XD 
is totally equivalent to a more sophisticated one based on 
counting the 
the number of particles in the largest cluster of solid-like particles. 
Solid-like particles can be identified
by means of a local-bond 
order parameter based on the local coordination of the particles  \cite{JCP_1996_104_09932}. The specific
parameters we use in this work are those given in Ref. \cite{peterRoySoc2009}. 
The largest cluster of solid-like particles corresponds to the crystal slab induced by the mold at coexistence conditions
(e. g., the 6-7 crystal
planes that can be seen around the mold in Fig. \ref{sketch}, bottom). 
In Fig. \ref{widthchoicei_app} we show the equivalent to Fig. \ref{widthchoice} 
but using the number of particles in the biggest cluster of solid-like particles, $n_s$, instead of $\xi$ as the parameter
to follow the formation of the crystal. By comparing both figures it is clear that
the simple parameter $\xi$ provides the same information as the more sophisticated $n_s$. 

\section{Free energy profiles.}
\label{fep}
\begin{figure}
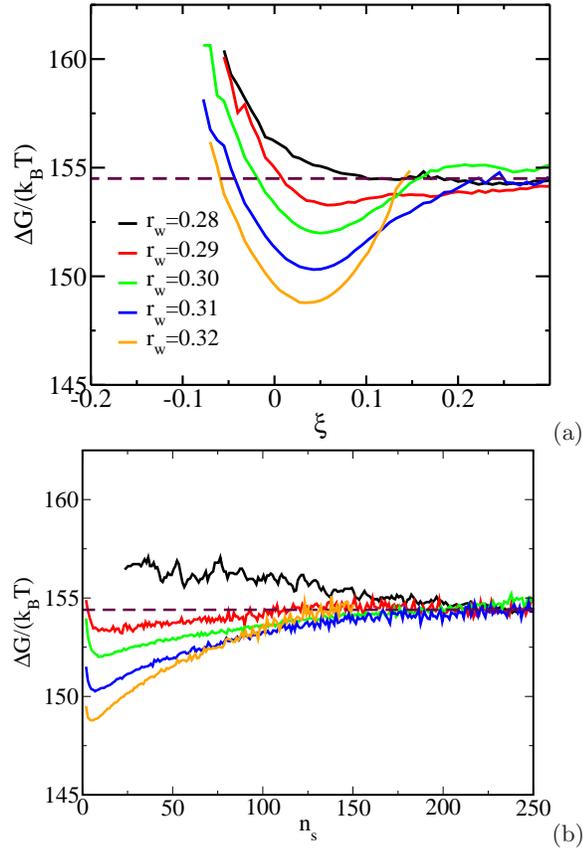
 \centering
\includegraphics[clip,width=0.40\textwidth]{FIG13A.eps}(a)
\includegraphics[clip,width=0.40\textwidth]{FIG13B.eps}(b)
\caption{Free energy profile as a function of XD for the 111 plane of the LJ
system as measured by
$\xi$, (a), and by $n_s$, (b) for different well radii (as indicated in the legend in
$\sigma$ units).}\label{deltagprof}
\end{figure}

By analysing the trajectories where XD is monitored (e.g. Figs. \ref{widthchoice} or \ref{widthchoicei_app}) 
the free energy profile along the XD coordinate 
can be estimated as:
\begin{equation}
\Delta G/(k_BT)=-\ln P(XD)  + constant
\label{dgp}
\end{equation}
where $P(XD)$ is the probability that the system takes a certain value of the 
order parameter, XD. Then, by simply making a histogram of XD for all trajectories performed 
for a given well radius it is possible to get an estimate of the corresponding free energy profile. 
In Fig. \ref{deltagprof} we show the free energy profile thus 
calculated for the 111 plane of the LJ system using both $\xi$ and $n_s$ as measures for XD. 
The conclusions that can be drawn by examining either order parameter are the same:
for $r_w \ge 0.29 \sigma$ there is a minimum whereas for $r_w \le 0.28 \sigma$ there is not.
Consequently, we set the optimal radius $r_w^o$ for this system to $0.285\pm0.005\sigma$.

Eq. \ref{dgp} does not give absolute 
free energies (there is a missing constant) but allows to determine whether there is a minimum 
present or not. In any case, in order to compare all curves in the same free energy scale
we have shifted each minimum to the work needed to fill
the wells for the corresponding well radius (calculated by thermodynamic integration via Eq.
\ref{tie}). For the cases where the minimum is absent ($r_w = 0.28$) we 
have shifted the plateau of the curves to the work needed to fill a mold of wells with
$r_w=r_w^o=0.285\sigma$, given by the dashed horizontal line. The statistics of the free energy
given by Eq. \ref{dgp} are reasonably good when the system repeatedly samples configurations 
in the vicinity of a free energy minimum but become poor when it quickly moves along 
the free energy plateau. Therefore, one has to be cautious and restrict the use of Eq. \ref{dgp}  
to small values of XD.

With the degree of accuracy we got in the present study we were able to distinguish
the anisotropy between the 111 and 100 faces of the LJ system.
One could in principle try to further improve the accuracy by decreasing the range within which 
$r_w^o$ is enclosed (by launching trajectories for more values of $r_w$).
This is certainly a possibility
worth exploring. However, it may require a substantial amount of trajectories to detect a mininum
shallower than 1$k_BT$, which is the depth of the shallowest minimum we could probe (curve
corresponding to $r_w=0.29$ in Fig. \ref{deltagprof}).

\end{document}